\documentclass[twocolumn,showpacs,preprintnumbers,amsmath,amssymb,superscript address, revsymb,prb]{revtex4}

\usepackage[dvips]{graphicx}
\usepackage{mathptmx}
\usepackage{dcolumn}

\newcommand{\gw}{G_0W_0^{\mathrm{RPA}}}	
\newcommand{\gwg}{GW\Gamma}			
\newcommand{\gwk}{GWK_{\mathrm{xc}}}		
\newcommand{\gwgl}{G_0W_0\Gamma^{\mathrm{LDA}}}	
\newcommand{\gwr}{G_0W_0^{\mathrm{RPA}}}	
\newcommand{\gwl}{G_0W_0^{\mathrm{LDA}}}	
\newcommand{\kxc}{K_\mathrm{xc}}		
\newcommand{\etal}{\emph{et al. }}		
\newcommand{\gkwgl}{G_0^{\mathrm{KLI}}W_0\Gamma^{\mathrm{LDA}}}
\newcommand{\gkwr}{G_0^{\mathrm{KLI}}W_0^{\mathrm{RPA}}} 
\newcommand{\gkwl}{G_0^{\mathrm{KLI}}W_0^{\mathrm{LDA}}} 
\newcommand{\gkw}{G_0^{\mathrm{KLI}}W_0}                 

\newcolumntype{.}{D{.}{.}{-3}}

\begin{document}
\title{Vertex Corrections in Localized and Extended Systems}

\author{Andrew J. Morris\footnote{Email: ajm255@cam.ac.uk. Present address: Theory of Condensed Matter Group, Cavendish Laboratory, University of Cambridge, J. J. Thomson Avenue, Cambridge CB3 0HE, United Kingdom.}}  
\affiliation{Department of Physics, University of York, Heslington, York
         YO10 5DD, United Kingdom}

\author{Martin Stankovski}
\affiliation{Department of Physics, University of York, Heslington, York
         YO10 5DD, United Kingdom}

\author{Kris T. Delaney\footnote{Present address: Materials Research Laboratory, University of California, Santa Barbara, CA 93106-5121, USA.}}
\affiliation{Department of Physics, University of Illinois at Urbana-Champaign, Urbana, IL 61801, USA}

\author{Patrick Rinke}
\affiliation{Fritz-Haber-Institut der Max-Planck-Gesellschaft, Faradayweg 4-6, D-14195 Berlin-Dahlem, Germany}

\author{P. Garc\'{i}a-Gonz\'{a}lez}
\affiliation{Departmento de F\'{i}sica Fundamental, U.N.E.D., Apartado 60141, E-28080 Madrid, Spain} 

\author{R. W. Godby}
\affiliation{Department of Physics, University of York, Heslington, York
         YO10 5DD, United Kingdom}

\date{\today{}}

\begin{abstract}
Within many-body perturbation theory we apply vertex corrections to various closed-shell atoms and to jellium, using a local approximation for the vertex consistent with starting the many-body perturbation theory from a DFT-LDA Green's function.
The vertex appears in two places -- in the screened Coulomb interaction, $W$, and in the self-energy, $\Sigma$ -- and we obtain a systematic discrimination of these two effects by turning the vertex in $\Sigma$ on and off.  We also make comparisons to standard $GW$ results within the usual random-phase approximation (RPA), which omits the vertex from both.
When a vertex is included for closed-shell atoms, both ground-state and excited-state properties demonstrate little improvement over standard $GW$.  For jellium we observe marked improvement in the quasiparticle band width when the vertex is included only in $W$, whereas turning on the vertex in $\Sigma$ leads to an unphysical quasiparticle dispersion and work function.
A simple analysis suggests why implementation of the vertex only in $W$ is a valid way to improve quasiparticle energy calculations, while the vertex in $\Sigma$ is unphysical, and points the way to development of improved vertices for {\it ab initio} electronic structure calculations. 
\end{abstract}
\pacs{71.45.Gm, 31.25.Eb, 31.25.Jf, 71.10.Ca}
\maketitle

\section{Introduction}
Many-body perturbation theory (MBPT) is a leading method for computing excited-state electronic  properties in solid-state physics~\cite{Aulbur2000, Aryasetiawan1997, H+L1969}. Within many-body perturbation theory, Hedin's $GW$ method~\cite{Hedin1965} is the most widely used approximation for the self-energy, $\Sigma$. The exact one-body Green's function, $G$ (which contains information about ground and excited-state properties of the system) can be written, using a Dyson equation, in terms of a suitable Green's function of a ``zeroth-order'' system of non-interacting electrons, $G_0$ (constructed from that system's one-particle wavefunctions and energies), and the self-energy operator $\Sigma$.  The approximation is defined by the choice of zeroth-order system and by the expression (typically a diagrammatic expansion in terms of $G_0$) used to approximate $\Sigma$. The self energy, $\Sigma$, contains all the information of many-body interactions in the system and can be obtained by using Hedin's set of coupled equations:
\begin{equation}
\label{Hedin_I}
{\Sigma}(12)=i{\int}W(1^{+}3)G(14){\Gamma}(42;3)\,d(34)
\end{equation}
\begin{equation}
\label{Hedin_II}
W(12)=v(12)+{\int}W(13)P(34)v(42)\,d(34)
\end{equation}
\begin{equation}
\label{Hedin_III}
P(12)=-i{\int}G(23)G(42){\Gamma}(34;1)\,d(34)
\end{equation}
\begin{align}
\label{Hedin_IV}
\nonumber
{\Gamma}(12;3)=&\delta(12)\delta(13)\\
&+{\int}\frac{{\delta}{\Sigma}(12)}{{\delta}G(45)}G(46)G(75){\Gamma}(67;3)\,d(4567),
\end{align}
and the Dyson equation, where $P$ is the polarizability, $W$ the screened and $v$ the unscreened Coulomb interaction and $\Gamma$ the vertex function.The notation $1\equiv({\bf x}_1,\sigma_1,t_1)$ is used to denote space, spin and time variables and the integral sign stands for summation or integration of all of these where appropriate. ($1^{+}$ denotes $t_1+\eta$ where $\eta$ is a positive infinitesimal in the time argument).  Atomic units are used in all equations throughout this paper. These are four coupled integro-differential equations where the most complicated term is the \emph{vertex}, $\Gamma$, which contains a functional derivative and hence, in general, cannot be evaluated numerically. The vertex is the usual target of simplification for an approximate scheme. 

The widely used $GW$ approximation is derived with the Hartree method as a starting point, and hence has a rigorous foundation only when started from a non-interacting Green's function, $G_0$, made from eigenstates of the Hartree Hamiltonian. This is because the initial self-energy, $\Sigma_{0} = 0$ and the vertex function is correspondingly set to ${\Gamma}(12;3)=\delta(12)\delta(13)$ since ${{\delta}{\Sigma}(12)}/{{\delta}G(45)}=0$.

Solving Hedin's equations with the vertex fixed in this expression yields the so-called self-consistent $GW$ approximation. In this approach, the
self-energy operator is formed from a product of a Green's function and a screened Coulomb interaction, where the Green's function used is consistent with that returned by Dyson's equation. Since the self-energy depends on $G$, this procedure should be carried out self-consistently, beginning with $G=G_0$.

In practice, it is customary to use the first iteration only, often called $\gw$, to approximate the self-energy operator. Here, $W_0^{\mathrm{RPA}}$ is perhaps the simplest  possible screened interaction, which involves an infinite geometric series over non-interacting electron-hole pair excitations as in the usual definition of the RPA.
It is important to make this one iteration as accurate as possible, so an initial $G_0$ calculated using Kohn-Sham density-functional theory in the local-density approximation (DFT-LDA) is normally used. This choice of $G_0$ generally produces much more accurate results for quasiparticle energies (the correct electron addition and removal energies, in contrast to the DFT-LDA eigenvalues\cite{Godby1986}). However, because this choice of $G_0$ corresponds to a non-zero $\Sigma_0$, there is no longer a theoretical justification for the usual practise of setting the vertex to a product of delta functions and different choices for the exchange-correlation functional may lead to different Green's functions \cite{Rinke_et_al, WillsPaper}.

Using the static exchange-correlation kernel, $\kxc$, (which is the functional derivative of the DFT exchange-correlation potential, $V_{\mathrm{xc}}$, with respect to density, $n$) Del Sole \etal\cite{DelSole}  demonstrated how $\gw$ may be modified with a vertex function to make $\Sigma$ consistent with the DFT-LDA starting point. They added the contribution of the vertex into both the self-energy, $\Sigma$, and the polarization, $P$. The result is a self-energy of the form $\gwgl$\,\cite{Notation}. The $\gwl$ approximation is obtained when the vertex function is included in $P$ only. As commented by Hybertsen and Louie\cite{Hybertsen1986} and Del Sole \etal, both these results take the {\it form} of $GW$, but with $W$ representing the Coulomb interaction screened by respectively the test-charge-electron dielectric function and the test-charge-test-charge dielectric function, in each with electronic exchange and correlation included through the time-dependent adiabatic LDA (TDLDA).

Del Sole \etal found that $\gwgl$ yields final results almost equal to those of $\gw$ for the band gap of crystalline silicon and that the equivalent results from $\gwl$ were worse when compared to $\gw$. It should perhaps be mentioned that the inclusion of other types of vertex corrections have been studied before as well, most notably corrections based on various approximations of a second iteration of Hedins equations, starting with $\gw$\cite{Arno1998,Bechstedt1997}. However, these have usually been applied with initial Kohn-Sham (KS) Green's functions, which are still not theoretically consistent with that starting point. The correct theoretical treatment of a second-iteration vertex from KS Green's functions is quite complicated and still absent in the literature. 

The purpose of the present work is to make a systematic study, for both localized and extended systems, of a simple \emph{ab initio} vertex correction whose form is determined by the starting approximation for the self-energy ($\Sigma_0=V_{\mathrm{xc}}$ for DFT-LDA). Related vertex corrections, including others derived from $\kxc$, have been investigated in earlier work. For example, Northrup \etal\cite{Northrup1987} used LDA bulk calculations as a starting point and a plasmon-pole calculation of the response function in conjunction with a $\gwl$-like vertex correction in the screened interaction. They found a narrowing of the band widths of Na, Li and K\cite{Surh1988} in agreement with the experiments of Jensen and Plummer\cite{Jensen1985} who had noted that the experimental band width was significantly narrowed ($\sim23\,\%$) compared to the free-electron result. Hedin's $\gw$~\cite{Hedin1965} calculations only gave a narrowing of about $10\,\%$ for an homogeneous electron gas of the same mean density, indicating a large impact of further many-body effects. This led to additional experimental and theoretical investigations\cite{Zhu1986,Shung1987,Lyo1988,Itchkawitz1990,Yasuhara1991,Frota1992} but the issue remains controversial \cite{Yasuhara1999,Takada2001,Ku2000,Yasuhara2000}.

For individual atoms, $GW$ quasiparticle properties have been investigated previously by Shirley and Martin\cite{Shirley1993} (including an exchange-only vertex) and, more recently, total energy studies on atoms and molecules using the variational functionals of Luttinger and Ward\cite{Luttinger1960} have been performed by Dahlen \etal\cite{Dahlen2004,Dahlen2006}, Stan \etal\cite{Stan2006} and Verdonck \etal\cite{Verdonck2006}. These studies have shown that $\gw$ in general gives quasiparticle properties which are much improved over DFT and Hartree-Fock methods and that, when calculated self-consistently, $GW$ also provides reasonably good total energies for atoms (with differences versus highly accurate reference methods being on the order of tens of $mHa$ per electron). To its merit, self-consistent $GW$ is also a conserving approximation in the Baym-Kadanoff\cite{Baym1961} sense. However, non-self consistent total energies in $\gw$ are noticeably less accurate. Conversely, the good agreement between the quasiparticle energies and experiment is destroyed when performing self-consistent calculations.

The answer to why this happens must, by definition, lie with the only approximated quantity, the vertex correction. This study is meant to address the need for a precise (including a full treatment in frequency) comparative study of the vertex corrections proposed by Del Sole \etal for localized and extended systems within $G_0W_0$.

\section{Method}
Hybertsen and Louie\cite{Hybertsen1986} comment that it is possible to start a $GW$ calculation from an initial self-energy, $\Sigma_{0}(12) = \delta(12)V_{\mathrm{xc}}(1)$. This approach gives a theoretical basis for beginning a $G_0W_0$ calculation from DFT-LDA orbitals. Adopting this idea, we see from Eq.\,(\ref{Hedin_IV}) that the second term is now non-zero, unlike in the $GW$ approximation.   Since the electron density is $n(1)= - iG(11^{+})$ then,
\begin{align}
\frac{\delta\Sigma(12)}{\delta G(45)}& =\frac{\delta \Sigma(12)}{\delta n(4)}\frac{\delta n(4)}{\delta G(45)} \\
& = -i\frac{\delta V^{\mathrm{LDA}}_{\mathrm{xc}}(1)}{\delta n(4)} = -i K^{\mathrm{LDA}}_{\mathrm{xc}}(1)
\label{Vertex}
\end{align}
where delta-functions are to be understood in all other variables.
In an appendix Del Sole \etal\cite{DelSole} show how to add this approximate vertex to both $W$ and $\Sigma$, and into $W$ only, by forming two different effective $W$s.  Our method follows that of Del Sole \etal\cite{DelSole} by modifying the dielectric function, $\epsilon$ from its form in the RPA. The screened Coulomb interaction in MBPT is written as
\begin{equation}
W = \epsilon^{-1}v,
\end{equation}
where $\epsilon^{-1}$ is the inverse dielectric function. We use the full polarization without recourse to plasmon-pole models. The random phase approximation (RPA) dielectric function is
\begin{equation}
\epsilon=1-v\chi^{0}.
\end{equation}
Del Sole \etal show that adding the form of the vertex from Eq.\,(\ref{Vertex}) into both $\Sigma$ and $W$ modifies the RPA dielectric function to,
\begin{equation}
\tilde{\epsilon}=1-\left(v+\kxc\right)\chi^{0},
\end{equation}
which leads to the introduction of an effective screened Coulomb interaction $\tilde{W}$. This is trivial to implement into a $GW$ computer code as it requires a simple matrix addition, once $\kxc$ is calculated.  The result of this modification is that  $\tilde{W}$ contains not only the screened Coulomb interaction but also an exchange-correlation potential. We shall refer to this method as $\gwgl$ as we have added the correct DFT-LDA vertex to the $GW$ method, hence the method is a one-iteration $GW\Gamma$ ($G_0W_0\Gamma_0$) calculation beginning with a DFT-LDA Green's function.

An alternative choice for the effective dielectric function,
\begin{equation}
\tilde{\epsilon}=1-\left(1-\kxc\chi^{0}\right)^{-1}v\chi^{0}
\end{equation}
corresponds to adding $\kxc$ into $W$ only. We term this method $\gwl$ as the LDA vertex is inserted into the screened Coulomb interaction, $W$, only. This is equivalent to the one-iteration $GW$ approximation, $G_0W_0$ but with $W$ calculated using the adiabatic LDA.

The vertices presented are thus dynamical, i.e. frequency dependent, due to the inclusion of $\chi^{0}$, and must include the  excitonic effects of the corresponding jellium due to the appearance of $\kxc$. Another way of looking at it is that this corresponds to a treatment beyond $G_0W_0$ where {\it at the level of the vertex corrections}, the system is modelled by the homogenous electron gas. It is not likely, however, that these methods would be able to capture any satellite structure beyond that provided by $\kxc$ as the calculations are non-self-consistent.

\section{Computational Approach}
The quasiparticle energies, $\epsilon_i$ and wavefunctions, $\psi_i$, are formally the solution of the quasiparticle equation,
\begin{align}
\label{qpe:generic}
\nonumber
\left\{ { - {\textstyle{1 \over 2}}\nabla ^2  + V_{{\rm ext}} ({\bf r}) + V_{\rm H} ({\bf r})} \right\}
  \:\psi_{i}({\bf r}) + &
  \int \Sigma({\bf r},{\bf r'};\epsilon_{i})  \psi_{i}({\bf r'})\, d{\bf r'}  \\
  &=\epsilon_{i} \psi_{i}({\bf r}). 
\end{align}
where $V_{{\rm ext}}$ and $V_{{\rm H }}$ are the external and Hartree potential, respectively.

In the case of a spherically symmetric system it is sufficient to describe all non-local operators in the $GW$ formalism  by two radial coordinates and one angular coordinate, $\theta$, that denotes the angle between the vectors ${\bf r}$ and ${\bf r'}$. The self-energy, $\Sigma$, then assumes the much simpler form,
\begin{equation}
  \label{eq:sigma_legexp}
  \Sigma(r ,r' ,\theta;\epsilon)=\sum_{l=0}^{\infty} \left[
                              \Sigma_l( r , r';\epsilon)\right]
                               P_l(\cos\theta),  
\end{equation}
where $P_l(\cos\theta)$ is a Legendre polynomial of order $l$.

The Legendre expansion coefficients of the self-energy, $\Sigma$, are calculated directly, thereby circumventing the need for a numerical treatment of the angular dependence. We use a real-space and imaginary time representation\cite{Rojas/Godby/Needs:1995} to calculate the self-energy from the non-interacting Green's function $G_0$. The self-energy on the real frequency axis, required for solving the quasiparticle equation, is obtained by means of analytic continuation\cite{Rojas/Godby/Needs:1995}. The current implementation has been successfully applied to jellium clusters\cite{ClusterImStates:2004} and light atoms\cite{Delaney/Garcia-Gonzalez/Rubio/Rinke/Godby:2004,WillsPaper}.

To obtain the quasiparticle energies and wavefunctions the quasiparticle equation (\ref{qpe:generic}) is fully diagonalized in the basis of the single particle orbitals of the non-interacting Kohn-Sham system. For localized systems the quasiparticle wavefunctions can differ noticeably from the wavefunctions of the non-interacting system or in certain cases even have a completely different character, as was demonstrated for image states in small metal clusters \cite{ClusterImStates:2004}.

Ground-state total energies were calculated using the Galitskii-Migdal formula\cite{Holm2000} transformed to an integral equation over imaginary frequency. This avoids analytic continuation of the self-energy, which would be unreliable for large frequencies.

For jellium, the homogenous electron gas, we solve Hedin's equations in wavevector and real-frequency space. This avoids analytic continuation and enables accurate and easy extraction of spectral properties. Again, we do not use plasmon-pole models, but the full frequency-dependent polarization.

\section{Total Energies}

\begin{table}[!h]
{\centering 
\begin{tabular}{c|d|d|d}
\hline\hline
Method     & \multicolumn{1}{c|}{He} & \multicolumn{1}{c|}{Be} & \multicolumn{1}{c}{Ne} \\
\hline
HF         & -1.4304^a               & -3.6433^b               & -12.8547^a    \\
DFT-LDA    & -1.4171                 & -3.6110                 & -12.8183      \\
$\gw$      & -1.4117(5)              & -3.5905(9)              & -12.777(1)    \\
$\gwl$     & -1.4120(2)              & -3.590(1)               & -12.775(15)   \\
$\gwgl$    & -1.3912(2)              & -3.573(1)               & -12.745(10)   \\
VMC        & -1.45176^a              & -3.66670^c              & -12.891(5)^a  \\
DMC        & -1.45186^a              & -3.66682^c              & -12.89231^a   \\
CI         & -1.45189^d              & -3.66684^d              & -12.89370^d   \\
\hline\hline
\end{tabular}\par}
\begin{flushleft}

$^a$ See reference~~\onlinecite{QMC} 

$^b$ See reference~~\onlinecite{BeHF}

$^c$ See reference~~\onlinecite{BeQMC}

$^d$ See reference~~\onlinecite{NewCI}

\end{flushleft}

\caption[Atomic Total Energy Methods]{Total energy data (Hartrees/electron). See Fig.\,\ref{Fig:AtomsTE} --- A comparison of various methods for total energy calculations. Hartree-Fock, Density-Functional Theory, one-iteration $GW$ ($\gw$), the two approximate vertex $GW$s, variational Monte Carlo, diffusion Monte Carlo and configuration interaction. (CI usually yields the most accurate estimate of the ground-state energies for localized systems.)}
\label{Tab:TEAtoms}
\end{table}

\begin{figure}
\includegraphics*[width=8cm]{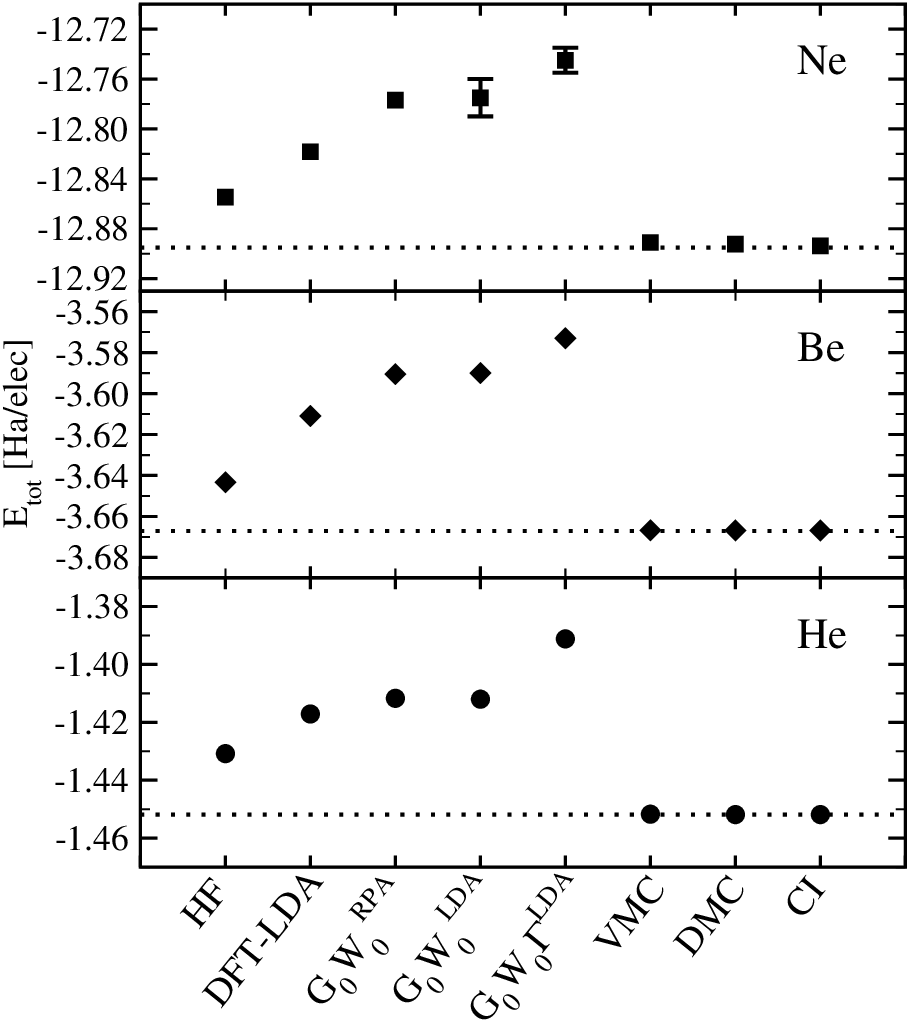}
\caption{Total energies of atoms. -- We compare a number of $GW$-based approaches to Hartree-Fock\cite{QMC}; DFT with an LDA exchange-correlation fucntional consistent with the $\kxc$ used; quantum Monte Carlo\cite{QMC} (VMC and DMC) and CI\cite{NewCI}.  The dotted line is the CI value and is there to guide the eye.  In all cases $\gwgl$ behaved poorly in comparison  to $\gw$, whereas $\gwl$ makes no improvement to $\gwr$. The MBPT methods are not as accurate as the computationally cheaper mean-field calculations but $\gw$ and $\gwl$ are the better of the three MBPT methods.}
\label{Fig:AtomsTE}
\end{figure}

\begin{table}[!h]
{\centering \begin{tabular}{c|c|c|c|c}
\hline\hline
$r_{\mathrm{s}}$ & \multicolumn{1}{c|}{2} & \multicolumn{1}{c|}{3} & \multicolumn{1}{c|}{4} & \multicolumn{1}{c}{5} \\
\hline
$\gw$           &$ -0.2826(3) $&$ -0.1967(1)    $&$ -0.1522(1)$ & $-0.1247(1)$ \\
$\gwl$          &$ -0.2857(4)$ &$ -0.2002(2)    $& $-0.1560(1)$ & $-0.1288(1) $\\
$\gwgl$         &$ -0.2525(2)$ &$ -0.1678(1)    $& $-0.1241(1)$ & $-0.0972(1) $\\
$GW~^a$         &$ -0.2727(5)$ &   n/a           & $-0.1450(5)$ & $-0.1185(5) $\\
QMC (DMC)~$^b$  &$ -0.2742(1)$ &$ -0.1902(1)   $ & $-0.1464(1)$ & $-0.1202(1) $  \\
\hline\hline
\end{tabular}\par}
\begin{flushleft}
$^a$ See reference~~\onlinecite{GarciaGonzales2001}

$^b$ See reference~~\onlinecite{Perdew1981}
\end{flushleft}

\caption{$\epsilon_{\mathrm{xc}}$ (Ha) --- The exchange correlation energy for jellium. The total energy per particle is given by $\epsilon=\frac{3}{5}\epsilon_{\mathrm{F}}+\epsilon_{\mathrm{xc}}$, where $\epsilon_{\mathrm{F}}=\frac{1}{2}k_{\mathrm{F}}^{2}=\frac{1}{2}\left(\frac{1}{\alpha{r_{\mathrm{s}}}}\right)^{2}$ and $\alpha=\left(\frac{4}{9\pi}\right)^{\frac{1}{3}}$. The energies under heading $GW$ are from self consistent calculations by Garc\'{\i}a-Gonz\'alez and  Godby\cite{GarciaGonzales2001} for reference.  $\gw$ is lower than the QMC energy by $\sim5\%$ on average. $\gwl$ is $\sim6\,\%$ lower and $\gwgl$ is $\sim10\,\%$ too high. The DMC values are evaluated by Perdew and  Zunger's\cite{Perdew1981} parametrization of Ceperly and Alder's DMC calculations.}
\label{Table:JelliumTE}
\end{table}

The MBPT total energy results are compared against configuration interaction (CI) and quantum Monte Carlo methods (variational Monte Carlo (VMC) and diffusion Monte Carlo (DMC)).  The CI and QMC family of methods usually yield the most accurate estimates of ground-state energies and are variationally bound, meaning that the lowest energy is the most accurate. 

The $G_0W_0$ result with $G_0$ constructed from DFT-LDA eigenstates, ($\gwr$) is in poor agreement with CI in all three cases. It is known that there is a large self-interaction error in the LDA, especially noticeable in smaller atoms.  Hartree-Fock, which is self-interaction free by construction, is more accurate than DFT-LDA.  Hence the self-interaction error is introduced via the LDA orbitals into the Green's function, $G_0^{\mathrm{LDA}}$, which gives rise to the $\gwr$ total energy's consistent poor agreement with CI.  (By way of illustration, using a $G_0$ from the superior KLI\cite{KLI}, an optimized effective potential that is formally free of self-interaction error for a two-electron system, greatly improves the DFT and $GW$ total energies. The $\gkwr$ results for He, Be and Ne are $-1.4550(3)$, $-3.6780(2)$ and $-12.843(1)$ respectively.)

For all three atoms the vertex in $W$ alone ($\gwl$) shows little difference to $\gwr$ (Table\,\ref{Tab:TEAtoms} and Fig.\,\ref{Fig:AtomsTE}), whereas $\gwgl$ raises the total energy with respect to $\gwr$. This change is due not to the LDA self-interaction but to the nature of the vertex. The result of adding the LDA vertex to $\gkw$ mirrors that of adding it to $G_0^{\mathrm{LDA}}W_0$. There is an increase of the total energy when the vertex is applied in both $W$ and $\Sigma$ ($\gkwgl$) but the vertex in $W$ only, ($\gkwl$), results in a similar total energy to $\gkwr$. (The $\gkwgl$ and $\gkwl$ for He are $-1.4235(10)$ and $-1.4475(5)$ respectively.) 

In jellium the trend is the same for all densities in the region from $r_{\mathrm{s}}=2$ to $5$ ($r_{\mathrm{s}}$ is the density parameter, where $r_{\mathrm{s}}=\left(\frac{3}{4\pi{n}}\right)^{1/3}$ and $n$ is the electron density in atomic units) as can be seen in Table\,\ref{Table:JelliumTE}. $\gwl$ lowers the total energy of jellium slightly as compared to $\gw$ and $\gwgl$ makes the energy too high. $\gw$ is on average $\sim5\%$ lower than the QMC result. $\gwgl$ is $\sim10\%$ too high and $\gwl$ $\sim6\%$ lower than the QMC result.

For jellium, neither method leads to a result more accurate than $\gw$. It is apparent, however, that the vertex added solely in the polarization has the minor effect of lowering the total energy. When the vertex is subsequently added into the self-energy there is a major positive shift in the total energy as seen in the atomic results as well. Self-consistent $GW$ calculations\cite{GarciaGonzales2001,Holm1998} for jellium show that the self-consistent total energy is about $4-5\%$ higher than the $\gw$ ones in the range of $r_{\mathrm{s}}=2$ to $5$ and the essentially exact QMC energies are about $0.5-1\%$ lower than the self-consistent $GW$ values. Assuming to a first approximation that the vertex corrections are independent and additive corrections to self-consistency, this would indicate that the $\gwgl$ energies would still be much too high, but the $\gwl$ energies might end up very close to the QMC results if self consistency is achieved, since they lower the $\gw$ energies to roughly the same extent as the difference between QMC and self-consistent $GW$ energies.

\section{Quasiparticle Energies}

\begin{table}[h]
{\centering \begin{tabular}{c|d|d|d}
\hline\hline
Method      & \multicolumn{1}{c|}{He} & \multicolumn{1}{c|}{Be} & \multicolumn{1}{c}{Ne}\\
\hline
DFT-LDA     & -15.4877                & -5.5909      & -13.503\\
$\gw$       & -24.20(4)               & -9.24(2)     & -20.55(10)\\
$\gwl$      & -24.05(5)               & -9.25(3)     & -19.48(10)\\
$\gwgl$     & -22.5(1)                & -7.56(6)     & -18.85(5)\\
CI          & -24.5930^a              & -9.3226^a    & -21.6034^a\\
Experiment  & -24.587^b               & -9.3227^c    & -21.5645^d\\
\hline\hline
\end{tabular}\par}
\begin{flushleft}
$^a$ See reference~~\onlinecite{NewCI}.  

$^b$ See references~~\onlinecite{HeExp1, HeExp2}.

$^c$ See reference~~\onlinecite{BeExp2}.

$^d$ See reference~~\onlinecite{NeExp}.

\end{flushleft}

\caption[Atomic Quasiparticle Methods]{First ionisation energy (eV) - A comparison of various methods for quasiparticle energy calculations: DFT-LDA, $\gw$ and the two approximate GWs. \emph{CI} denotes the ionisation potential calculated from the difference in CI total energies and \emph{Experiment} the measured value.}
\label{Table:AtomsQPE}
\end{table}

\begin{figure}
\includegraphics*[width=8cm]{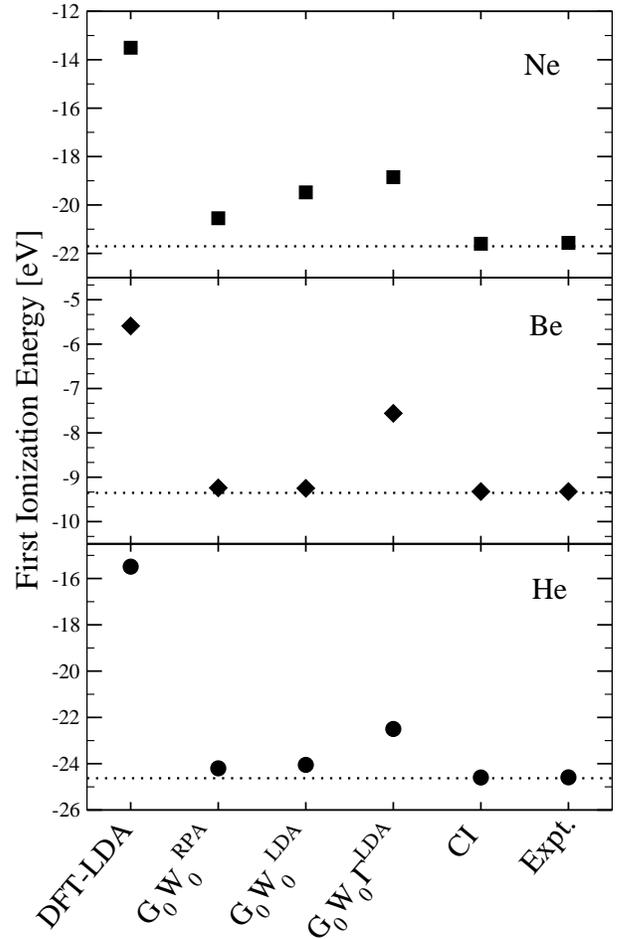}
\caption{QP energies of atoms - The first ionization energy is plotted. We compare the $GW$ based approaches to DFT-LDA; the experimental\cite{HeExp1, HeExp2,BeExp2} answer (Expt.) and the values calculated from the differences in CI\cite{NewCI} calculations (CI). The helium, beryllium and neon values are plotted with circles, diamonds and squares respectively. The dotted lines go through the CI value  and are there to guide the eye.  All of the $GW$ calculations are more accurate than the mean-field method. For helium and beryllium the $\gw$ and $\gwl$ methods give similar results. In all cases $\gwgl$ is shifted to a higher energy that $\gw$.  (This shift for $\gwgl$ was also found by Del Sole \etal\cite{DelSole} and Fleszar and Hanke\cite{Fleszar}.)}
\label{Fig:AtomsFIE}
\end{figure}

The quasiparticle energy corresponding to the first ionization energy \footnote{We compute quantities for the non-relativistic, all-electron Hamiltonian} is presented for helium, beryllium and neon in Fig.\,\ref{Fig:AtomsFIE}. The MBPT methods are consistently more accurate than DFT-LDA Kohn-Sham eigenvalues. However, again $\gwl$ is roughly equivalent to $\gw$ for helium and beryllium and in all cases $\gwgl$ causes an increase in quasiparticle energy, in agreement with Del Sole \etal\cite{DelSole}.  In general, $\gwgl$ worsens QP energies as compared to $\gw$. 

\begin{figure}
\includegraphics[width=8cm]{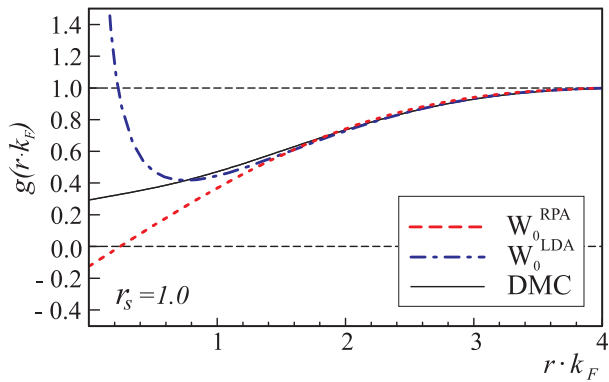}
\caption{The pair-correlation function --- evaluated at $r_{\mathrm{s}}=1.0$. The RPA goes negative for small $r$, which is a well-known unphysical behaviour of this approximation. Including the vertex makes the $W^{LDA}_0$ fit the (essentially exact) QMC\cite{GoriGiorgi2000} curve better, but then the p.-c. function goes too positive instead. The $W^{LDA}_0$ curve does not go to infinity when $r\rightarrow0$, it just goes to a very high value ($\sim2000$). The horizontal lines are there to guide the eye.}
\label{Fig:JelliumPCF}
\end{figure}

For jellium, different quantities are accessible at different stages of the iteration of Hedin's equations. The pair-correlation function $g(r)$ for example, can be obtained from the (isotropic) inverse dielectric function, $\epsilon^{-1}(k,\omega)$, by the integration
\begin{equation}
  \label{eq:paircorr1}
  g(r)=1+\frac{3}{2rk^{3}_{F}}\int_{0}^{\infty}dk\,k\sin(kr)\left[S(k)-1\right],
\end{equation}
where the static structure factor,
\begin{equation}
  \label{eq:paircorr2}
  S(k)=-\frac{k^2r^3_{\mathrm{s}}}{3}\int_{0}^{\infty}\frac{d\omega}{\pi}\Im\left[\epsilon^{-1}(k,\omega)\right].
\end{equation}
 $g(r)$ is shown in Fig.\,\ref{Fig:JelliumPCF} for $r_{\mathrm{s}}=1.0$. The RPA displays the well known failure to be positive definite for $r_{\mathrm{s}}\gtrsim0.78$. This is remedied by the local vertex, but the result appears to be an overcorrection (note that $\gwl$ and $\gwgl$ are equivalent at this stage since $\Sigma$ has not yet been calculated).

\begin{figure}
\includegraphics[width=8cm]{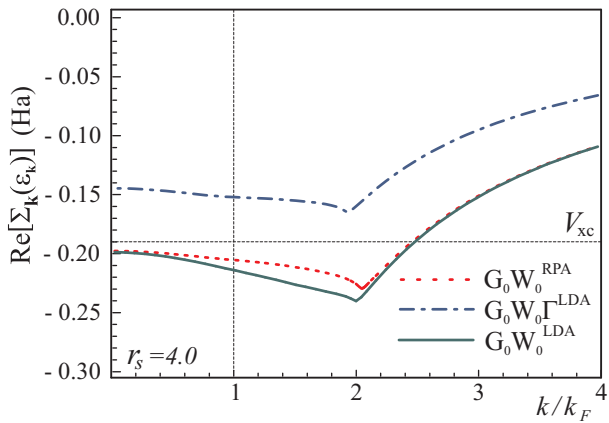}
\caption{Correction to the free-electron quasiparticle dispersion for $r_{\mathrm{s}}=4.0$--- Note the large absolute shift for $\gwgl$. The two intersecting straight lines are included to emphasize that none of the dispersions fulfill the condition $Re\left[\Sigma_{{\bf k}_{F}}(\epsilon_{F})\right] = V_{xc}$.}
\label{Fig:JelliumQPD}
\end{figure}

The tendency of $\gwgl$ to overshoot -- the reason for which, we will defer to the closing discussions -- is apparent in all subsequent results. Once $\Sigma$ has been calculated, the QP dispersion can be extracted. Presented in Fig.\,\ref{Fig:JelliumQPD} is the real part of the self-energy evaluated at the self-consistent eigenvalues, \emph{i.e.} the correction $Re\left[\Sigma_{{\bf k}}(\epsilon_{{\bf k}})\right]$ to the quasiparticle dispersion as found by the formula
\begin{equation}
\label{qp_eigenvals}
\epsilon_{{\bf k}}=\epsilon^{0}_{{\bf k}}+Re\left[\Sigma_{{\bf k}}(\epsilon_{{\bf k}})\right],
\end{equation}
where $\epsilon^{0}_{\bf k}$ is the non-interacting dispersion. Care has been taken to align the Fermi energy of non-interacting and interacting systems so that the Dyson equation is consistent\cite{Holm2000} and all quantities are calculated in real frequency. The self-consistent quasiparticle energy should be used when one has a self-consistent $\Sigma$, but for a $G_0W_0$ calculation there is still controversy about whether the self-consistent eigenvalues $\epsilon_{\bf k}$ or the zeroth-order eigenvalues $\epsilon^{0}_{\bf k}$ are best used as the argument of $\Sigma_{{\bf k}}$ in equation (\ref{qp_eigenvals})\cite{Aulbur2000,PrivateCom}. The self-consistent approach was chosen in this paper.

\begin{table}[!h]
{\centering \begin{tabular}{c|c|c|c|c}
\hline\hline
$r_{\mathrm{s}}$ &  $\gw$  & $\gwgl$ &   $\gwl$ &   Experiment   \\
\hline
(Al)  2.07       & 11.5445 & 11.6444 &  11.1814 & 10.60$\pm$0.10~$^a$   \\
(Li)  3.28       &  4.4644 &  4.4853 &   4.2129 & ~3.00$\pm$0.20~$^b$   \\
(Na)  3.96       &  2.9837 &  2.9889 &   2.7777 & ~2.65$\pm$0.05~$^c$   \\
(K)   4.96       &  1.8625 &  1.8579 &   1.7044 & ~1.60$\pm$0.05~$^d$   \\
(Rb)  5.23       &  1.6669 &  1.6610 &   1.5191 & ~1.70$\pm$0.20~$^e$   \\
(Cs)  5.63       &  1.4287 &  1.4215 &   1.2944 & ~1.35$\pm$0.20~$^e$   \\
\hline\hline
\end{tabular}\par}
\begin{flushleft}
$^a$ See reference~~\onlinecite{Levinson1983}. 

$^b$ See reference~~\onlinecite{Crisp1960}.

$^c$ See reference~~\onlinecite{Lyo1988}.

$^d$ See reference~~\onlinecite{Itchkawitz1990}.

$^e$ See reference~~\onlinecite{Smith1971}.
\end{flushleft}
\caption{Occupied band widths of jellium for different $r_{\mathrm{s}}$ (eV) --- evaluated at the self-consistent eigenenergy.}
\label{Table:JelliumBandwidth}
\end{table}

\begin{figure}
\includegraphics*[width=8cm]{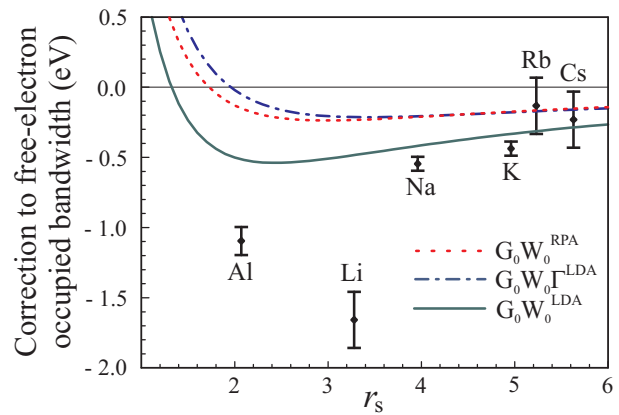}
\caption{Correction to the free-electron occupied band width --- Comparison with experiments. $\gwl$ has the most significant narrowing of the band in the relevant density-region. $\gwgl$ is not very different from $\gw$. Note that a jellium calculation does not include the contribution from the crystal lattice potential.} 
\label{Fig:JelliumBandwidth}
\end{figure}

The difference between the quasiparticle energies at $k=k_{\rm F}$ and $k=0$ is known as the band width, which therefore takes the form of the free-electron value ($k_{\rm F}^2/2)$ corrected by the change in Fig. \ref{Fig:JelliumQPD} between $k=k_{\rm F}$ and $k=0$.  This band width is shown in Table \ref{Table:JelliumBandwidth} and Fig. \ref{Fig:JelliumBandwidth}.
It consistently seems that vertex corrections applied in the screened Coulomb interaction \emph{only} give the best results. This is corroborated by the fact that this quasiparticle dispersion has a better band width and that the $\gwgl$ introduces little change to the band width. These results are in agreement with those of Mahan and Sernelius\cite{MahanAndS} obtained for a model Hubbard vertex.

\section{The chemical potential of jellium}

\begin{figure}
\includegraphics*[width=8cm]{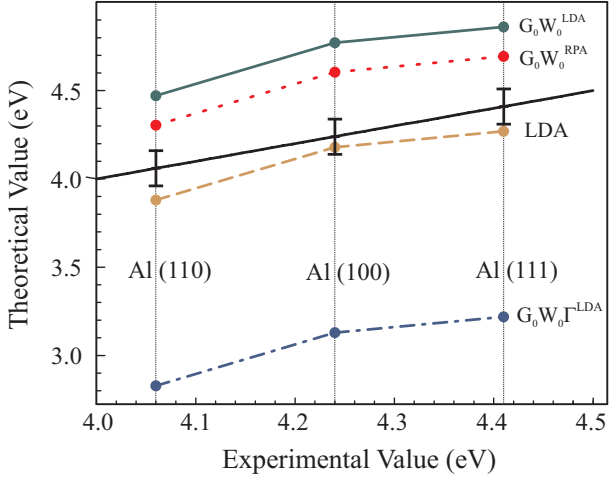}
\caption{(Color online) The workfunction of aluminium ($r_{\mathrm{s}}=2.07$) --- Compared to experiment. The LDA surface calculation corresponds to the exact result if jellium is used to model the bulk Fermi energy. $\gw$ and $\gwl$ are closer, while $\gwgl$ is much worse. The colored lines are there to guide the eye. The black line corresponds to perfect agreement with experiment (error bars indicate experimental uncertainty)}
\label{Fig:JelliumWkFn}
\end{figure}

To get another indication of whether the large absolute positive shift of the quasiparticle dispersion is physical, we compare with experimental work functions of Al (1\,0\,0), (1\,1\,0) and (1\,1\,1) surfaces (see Table \ref{Table:JelliumWkFn}). We assume that the electron density of the surface region, and therefore the electrostatic surface-dipole energy barrier, are well described by LDA calculations including the crystal lattice.  The work function, $\phi$, will, however, be sensitive to the quasiparticle bulk Fermi level, which we use here as a discriminator between self-energy approximations in the bulk.  Treating the bulk metal as jellium, we obtain a shift in the workfunction due to the new chemical potential,
\begin{equation}
\label{workfunction}
\phi=\phi^{\mathrm{LDA}}+\Delta\mu=\phi^{\mathrm{LDA}}-\left(Re\left[\Sigma_{{\bf k}_{\mathrm{F}}}(\epsilon_{\mathrm{F}})\right]-V_{\mathrm{xc}}\right).
\end{equation}
where $\Delta\mu$ is the correction due to the shift of the bulk Fermi energy for a $GW$ jellium calculation. The LDA workfunction is defined as the shift between the vacuum potential, $\phi_{\mathrm{vac}}$, and the chemical potential from the LDA surface calculation, $\mu^{\mathrm{LDA}}$. Since the exact self-energy for jellium must fulfill the condition
\begin{equation}
\label{sigma_exact}
Re\left[\Sigma_{{\bf k}_{\mathrm{F}}}(\epsilon_{\mathrm{F}})\right]=V_{\mathrm{xc}},
\end{equation}
we see that the LDA (taken from highly accurate QMC calculations) corresponds to the exact result if one assumes that the bulk is accurately modelled by jellium. Comparing with Table \ref{Table:JelliumWkFn} and Fig.\,\ref{Fig:JelliumWkFn} we see that $\gw$ is closest to the exact result and $\gwl$ is slightly further away, while $\gwgl$ is even worse.

\begin{table}[!h]
{\centering \begin{tabular}{c|c||c|c|c|c}
\hline\hline
$\Phi$ Al     &  Exp. (eV)$^a$ & LDA$^b$ & ${\gw}$  &  ${\gwgl}$ & ${\gwl}$   \\
\hline
(1\,0\,0)     &    4.41        & -0.14   &   0.28   &   -1.19    &   0.45     \\
(1\,1\,0)     &    4.06        & -0.18   &   0.24   &   -1.23    &   0.41     \\
(1\,1\,1)     &    4.24        & -0.06   &   0.36   &   -1.11    &   0.53     \\
\hline\hline
\end{tabular}\par}
\begin{flushleft}
$^a$ See reference~~\onlinecite{Michaelson1977}. 

$^b$ See reference~~\onlinecite{Serena1988}.
\end{flushleft}

\caption{The workfunction of aluminium ($r_{\mathrm{s}}=2.07$) --- Compared to experiment. The last four columns show the deviation from the experimental value. The LDA surface calculation corresponds to the exact result if jellium is used to model the bulk fermi energy. $\gw$ and $\gwl$ are closer, while $\gwgl$ is much worse}
\label{Table:JelliumWkFn}
\end{table}

This leads us to conclude that $\gwgl$ is unphysical in the sense that a vertex correction derived from a self-energy approximation with a completely local dependence on the density (like the LDA) will have pathological features. This is most probably due to improper behaviour of the spectral function of the screened interaction, as is demonstrated in the final section of this paper.

\section{Discussion and Conclusions}

We have presented $\gwl$ and $\gwgl$ calculations for isolated atoms and jellium.  We see that $\gwgl$ worsens results in all cases compared to the common $\gw$ approximation.

A proper \emph{ab initio} vertex correction for calculations on an arbitrary system should be derived from the starting approximation for the self-energy. In this work we have shown that in practice, vertex corrections derived from a \emph{local} density approximation to the self-energy (like the LDA) are pathological when applied to both the self-energy and the screened interaction. The work function of aluminium was used to confirm that the value of the chemical potential in $\gwgl$ is far from the correct result.

An indication of why a local correction in both $W$ and $\Sigma$ performs so poorly has been discussed previously by Hindgren and Almbladh\cite{Hindgren1997} and investigated in excitonic effects on wide-bandgap semiconductors by Marini \etal\cite{Marini1,Marini2}. Both types of vertex corrections lead to a modified screened interaction $\tilde{W} = \tilde{\epsilon}^{-1}v$. The spectral function of this, which is required to be positive semidefinite for $\omega\leq0$ and negative semidefinite otherwise, is given by
\begin{align}
\label{spectral_function}
\nonumber
B_{\bf q}(\omega)=-\frac{\mathrm{sgn}(\omega)}{\pi}\mathrm{Im}\left[\tilde{W}_{\bf q}(\omega)\right] \\
                 =\frac{\mathrm{sgn}(\omega)}{\pi}\frac{\mathrm{Im}\left[\tilde{\epsilon}_{\bf q}(\omega)\right]}{\left|\tilde{\epsilon}_{\bf q}(\omega)\right|^{2}}v_{\bf q},
\end{align}
so it inherits whatever properties of definiteness the imaginary part of the dielectric function has. Now for $\gwgl$ this is given by,
\begin{equation}
\label{GWGL_Im_diel}
\mathrm{Im}\left[\tilde{\epsilon}_{\bf q}(\omega)\right]=-(v_{\bf q}+K_{xc})\mathrm{Im}\left[\chi^{0}_{\bf q}(\omega)\right].
\end{equation}
Since the $RPA$ response function, $\chi_{0}$, will have the correct analytical properties by construction, this expression will obviously change sign whenever $K_{xc}$ - which is strictly negative for all densities and a negative constant for jellium - is larger in magnitude than $v_{\bf q}$, which decays as $1/q^2$. This will thus lead to a spectral function with the wrong sign, which is entirely unphysical. For jellium, isolated atoms, or any sparse enough condensed state, this is guaranteed to happen, because $K_{xc}\rightarrow-\infty$ for low densities.  Inspection of the dielectric function in $\gwl$,
\begin{equation}
\label{GWL_Im_diel}
\mathrm{Im}\left[\tilde{\epsilon}_{\bf q}(\omega)\right]=-\frac{v_{\bf q}\mathrm{Im}\left[\chi^{0}_{\bf q}(\omega)\right]}{\left|1-K_{xc}\chi^{0}_{\bf q}(\omega)\right|^2},
\end{equation}
illustrates that it cannot suffer from the same pathology. Since Eq. (\ref{GWL_Im_diel}) ensures that the static structure factor has the correct behavior for both $\gwl$ and $\gwgl$, no conclusions can be drawn on the reason for the overly positive value of the pair-correlation function of jellium when $r\rightarrow0$, except that it must depend on the high $k$ behavior of the denominator. We note that none of the calculations have been carried out self-consistently; it is possible that the vertices presented here go some way to improve self-consistent $GW$ results\footnote{Recently Fleszar and Hanke\cite{Fleszar2} have also used the $\gwgl$ vertex and noted improvement in band gaps and the energy positions relative to the valence band minimum for the computationally difficult II$^{\mathrm{B}}$-VI semiconductors when the $\gwgl$ method is used in conjunction with $G$ and $W$ partially updated through one previous iteration of Hedin's equations.}.

One possibility of the failure of the LDA starting point with the inclusion of the theoretically consistent vertex is the self-interaction error the LDA orbitals carry with them. Any starting point with an inherent self-interaction error should lead to correcting terms in the diagrammatic expansion. It is possible that the first-order correction, like $\gwgl$, is not enough and higher order corrections must be applied. A vertex derived from a second iteration of Hedin's equations does indeed lead to further and more complicated diagrams than the equivalent vertex from a Hartree starting point. Unfortunately these diagrams are of prohibitive complexity for practical calculations. 

It is still not understood why a correction in $W$ only (in a TDDFT-like manner) seems to work as well as it does. There is a similarity here with the way that the Bethe-Salpeter equation (BSE) is usually applied for the calculation of optical spectra. There too it is well known that, in theory, inclusion of a screened interaction in electron-hole excitations should be accompanied by an inclusion of the double-exchange term in $\Sigma$ but this has been proven to worsen results. Recently, Tiago and Chelikowsky\cite{Tiago2006} have used a $\gwl$ vertex in conjunction with an efficient numerical implementation of the BSE for isolated molecules and have shown that the inclusion of a TDLDA vertex gives very good results over a wide range of structural configurations in excited states.

For atomic helium and beryllium $\gwl$ is very similar to the $\gw$ result but slightly worse in neon. While in jellium the band width is improved. Hence $\gwl$ may be a local and easily implementable way to improve quasiparticle results in extended systems.

Overall, vertices based on the local density clearly have their limitations, arriving in part from the wavevector independent character of $\kxc$.  It should be fruitful to explore vertices that incorporate non-local density-dependence and reflect the non-local character of the original self-energy operator.

\begin{acknowledgments}
The authors would like to thank Ulf von Barth, Carl-Olof Almbladh, Peter Bokes, Arno Schindlmayr, Matthieu Verstraete, Steven Tear and John Trail for helpful discussions. This research was supported in part by the European Union (contract NMP4-CT-2004--500198, ``Nanoquanta'' Network of Excellence), the Spanish MEC (project FIS2004-05035-C03-03) and the Ram\'{o}n y Cajal Program (PGG).
\end{acknowledgments}


\begin{thebibliography}{10}
\bibitem{Notation}
We have modified the notation of Del Sole \etal to clarify the precise nature of the different approximate vertices: we use $\gwgl$ in place of $\gwg$, and $\gwl$ in place of $\gwk$.
\end{thebibliography}
\end{document}